\documentclass[aps,preprint]{revtex4}%
\usepackage{amsfonts}
\usepackage{amsmath}
\usepackage{amssymb}
\usepackage{graphicx}%
\providecommand{\U}[1]{\protect\rule{.1in}{.1in}}

\begin{document}
\title{Muon-spin-rotation study of magnetism in $\mathbf{Na_{x}CoO_{2}}$ single
crystals with$\ 0.78\leq x\leq0.97$}
\author{C. Bernhard$^{1,2}$, Ch. Niedermayer$^{3}$, A. Drew$^{1}$, G. Khaliullin$^{2}%
$, S. Bayrakci$^{2}$, J. Strempfer$^{2},$ R.K. Kremer$^{2}$, D.P. Chen$^{2}$,
C.T. Lin$^{2}$, and B. Keimer$^{2}$}
\affiliation{1) University of Fribourg, Department of Physics and Fribourg Center for
Nanomaterials (FriMat), Chemin du Musee 3, CH-1700 Fribourg, Switzerland.}
\affiliation{2) Max-Planck-Institute for Solid State Research, Heisenbergstrasse 1,D-70569
Stuttgart, Germany.}
\affiliation{3) Paul-Scherrer-Institut, CH-5232 Villigen PSI, Switzerland.}


\begin{abstract}
By muon spin rotation we investigated the magnetic properties of a series of
highly Na doped Na$_{x}$CoO$_{2}$ single crystals with 0.78(1)$\leq x\leq
$0.97(1). Our data provide evidence for an intrinsically inhomogeneous
magnetic state which can be described in tems of hole doping (Na vacancy)
induced magnetic clusters that percolate at 1-$x\gtrsim0.04$ until they yield
a bulk magnetic state near $x=$0.78. Evidence for a strong (likely
geometrical) frustration of the magnetic order is obtained from the anomalous
doping dependence of the spin fluctuation rate (above the ordering
temperature) which is strongly enhanced at $x=$0.78 as compared to $x=$0.97.

\end{abstract}
\date{\today}
\maketitle

The discovery of superconductivity with T$_{c}\leq$5 K in the hydrated layered
cobaltate $\mathrm{Na_{0.35}CoO_{2}\ast1.3\,H_{2}O}$ \cite{Takada03} has
renewed the interest in the physical properties of the parent compound
Na$_{x}$CoO$_{2}.$ Its rich phase diagram includes several phases with
extraordinary electromagnetic and thermoelectric properties
\cite{Foo03,Huang04,Lee06}. One attractive feature is the triangular
coordination of Co which may favor geometrical frustration and unconventional
ground states \cite{Tanaka03,Baskaran03}. Another curious aspect concerns the
finding that the anomalous electromagnetic properties are not confined to the
Mott-insulator regime at low $x$, but also occur in the vicinity of the
anticipated band-insulator at $x$=1. Examples are the occurrence of magnetic
order at $x\geq0.75$ \cite{Sugiyama03,Bayrakci03,Bayrakci05,Helme05,Mendels05}%
, signatures of strong spin-charge coupling \cite{Bernhard04} and giant
thermoelectric effects \cite{Lee06}. The nature of the underlying magnetic
state is still debated. While initial data were interpreted in terms of an
incommensurate spin-density-wave (SDW) \cite{Sugiyama03}, recent neutron- and
muon-spin-rotation ($\mu$SR) studies on single crystals
\cite{Bayrakci03,Bayrakci05,Helme05,Mendels05} yield commensurate A-type
antiferromagnetic (AF) order and a weakly anisotropic magnetic exchange
coupling constant that contrasts with the large Fermi-surface anisotropy.
Besides, an interpretation in terms of localized Co moments in the
low-spin-state is also questionable since the long range magnetic order would
occur in a strongly diluted state with only 1-$x$ magnetic Co$^{4+}$ ions
($^{5}$t$_{2g}$, S=1/2) and $x$ non magnetic Co$^{3+}$ ions ($^{6}$t$_{2g}$, S=0).

Here we present muon spin rotation ($\mu$SR) and dc magnetization measurements
which provide new insight into the evolution of the magnetic properties of
highly Na doped Na$_{x}$CoO$_{2}$ single crystals with 0.78(1)$\leq x\leq
$0.97(1). Our data provide evidence for hole doping (or Na vacancy) induced
magnetic clusters which remain isolated at 1-$x\lesssim$0.04 but percolate at
1-$x$%
$>$%
0.04 until a homogeneous magnetic state is achieved near 1-$x\approx$0.25. We
also find that this long-range ordered magnetic state exhibits signatures of
strong (likely geometrical) frustration.

The Na$_{x}$CoO$_{2}$ crystals have been cleaved from ingots that were grown
with an optical floating-zone furnace \cite{Lin05}. The average Na:Co ratio of
our crystals has been carefully determined by inductively coupled plasma
spectroscopy (ICPS) and by neutron activation analysis with an accuracy of
about 1\%. The crystals were also characterized by electrical transport, dc
magnetometry, and single crystal x-ray diffraction. The latter suggest that
the $x$=0.78 crystals are single phase (likely $\alpha%
\acute{}%
$), while the ones at $x$=0.87, 0.92 and 0.97 contain mixed $\alpha$- and
$\alpha%
\acute{}%
$- phases, similar as was previously reported \cite{Huang04,Lee06}. For the
latter, the intensity of the $\alpha%
\acute{}%
$- phase peaks was found to exhibit a strong variation, even for crystals with
a similar bulk Na content (according ICPS\ and neutron activation analysis).
Accordingly, we suspect as outlined below that the x-ray data my be strongly
affected by Na deficient surface layers. Finally, the $\mu$SR measurements
were performed with the GPS setup at the $\pi$m3 beamline of the
Paul-Scherrer-Institut (PSI) in\ Villigen, Switzerland, which provides 100\%
spin-polarized muons. Details concerning the technique can be found in
\cite{Lee98}. Here we only note that zero-field $\mu$SR technique provides
information about the local scale distribution of the internal magnetic fields
of magnetic materials. In particular, it yields a reliable estimate of the
bulk volume fractions of the magnetic phases, even for the case of weak and
strongly disordered magnets.

First we discuss the $\mu$SR\ data of the Na$_{0.97}$CoO$_{2}$ crystal with
1-$x$=0.03(1) holes per Co (Figure 1). The zero-field (ZF) spectra between 5
and 300 K are shown in Fig. 1(a). The low-\textit{T} spectra evidently consist
of two distinct parts. Nearly 40\% of the signal depolarize rapidly while the
remaining 60\% exhibit a much slower relaxation. This suggests that the sample
is in a spatially inhomogeneous magnetic state with about 40\% of the volume
containing sizeable magnetic moments and around 60\% remaining non magnetic.
It is also evident that the correlations in the magnetic regions persist to
temperatures in excess of 200 K. The absence of oscillations in the fast
relaxing signal implies a broad distribution of the magnetic fields that
originates either from static but spatially disordered moments or else from
dynamical fluctuations. It is well known that a large longitudinal field (LF)
enables one to distinguish between these two cases \cite{Lee98}. It gives rise
to a strong reduction of the amplitude of the magnetic signal (so-called
\textquotedblright decoupling effect\textquotedblright) in the static case but
not in the dynamic one. Accordingly, the data at LF=1 kOe in Figs. 1b and 1c
establish that a transition from a static to a dynamic state takes place
between 15 and 30 K. The signatures of the corresponding spin freezing
transition are also apparent in the transverse field (TF) relaxation rate,
$\lambda^{TF}$, in Fig. 1d. The slow increase of $\lambda^{TF}$ between 300
and 50 K is consistent with a gradual decrease of the spin fluctuation rate,
while the following rise and the saturation below 20 K are characteristic of a
freezing transition at $T_{f}\approx$20 K. Furthermore, the transition width
and the large value of $\lambda_{0}^{TF}$ are characteristic for a glassy
state \cite{Zhang05}. This freezing transition is also evident in the dc
magnetic susceptibility in Fig. 2 due to a deviation from the Curie law below
50 K and a cusp around $T_{f}$=20 K, as marked by the arrow.

The inhomogeneous magnetic state for samples with $x\approx1$ was previously
noticed but it was discussed in terms of chemical inhomogeniety due to
inclusions of a Na deficient and thus magnetic phase \cite{Julien05} or even
of impurity phases \cite{Mendels05b}. The former interpretation was motivated
by x-ray diffraction data which signal the presence of two Na$_{x}$CoO$_{2}$
phases with different Na content \cite{Huang04,Lee06,Julien05,Mendels05b}.
However, in the following we argue that such a purely chemical scenario is not
sufficient to explain our data. Instead, we present evidence that the
formation of intrinsic, doping induced magnetic nanoclusters seems to play an
important role. For example, we investigated crystals from two growth batches
with $x=$0.97(1) out of which one also exhibited the x-ray signatures of a
secondary Na deficient $\alpha%
\acute{}%
$-phase, while the other according to x-rays was an almost pure $\alpha$-phase
(see inset of Fig. 1d). Despite of this substantial difference, our $\mu$SR
data yield virtually identical bulk magnetic properties for both crystals (Fig
1 shows our $\mu$SR data of the pure $\alpha$-phase crystals). The only
distinction concerns a minor difference in the magnetic volume fraction that
is well consistent with the 1\% uncertainty in the Na content. These
observations suggest that one needs to distinguish between a nanoscale phase
separation as probed by $\mu$SR and a macroscopic one as evidenced for example
with x-rays. While a strong correlation between the Na vacancies and the doped
holes which determine the electronic and magnetic properties of the CoO$_{2}$
layers is evident \cite{Roger07}, it remains unknown how the nanoscale
magnetic clusters are related to the macroscopic Na deficient regions that are
frequently observed in x-ray \cite{Julien05,Mendels05b} and neutron
diffraction \cite{Huang04}. We noticed however, that the x-ray signature of
the $\alpha%
\acute{}%
$-phase in highly Na doped crytstals appears to be time dependent, even on the
scale of weeks. To the contrary, we confirmed that the $\mu$SR data do not
exhibit any noticeable changes, not even after one year. Accordingly, we
suspect that the apparent instability of the crystal surface at ambient
conditions towards the formation of Na$_{2}$CO$_{3}$ and a subsequent
development of a Na deficient surface layer may play an important role here.
We note that this is especially relevant for polycrystalline samples on which
the majority of experiments has been performed
\cite{Huang04,Sugiyama03,Mendels05,Mendels05b}. In this context, we emphasize
that our crystals are mm sized and that $\mu$SR is a truly bulk sensitive
technique since the muon implantation depth is of the order of 100 $\mu$m.

Our combined $\mu$SR and dc magnetisation measurements are indeed consistent
with the point of view that the relevant length scale of the magnetic phase
separation at $x$=0.97 is on the order of nanometers. In the first place they
are suggestive of an antiferromagnetic spin coupling within the magnetic
clusters. Specifically, the dc magnetisation data in Fig. 2 yield a fairly
small value of the total Curie-moment is otherwise difficult to reconcile with
the large magnetic volume fraction and the sizeable magnetic Co moment that
emerges from the $\mu$SR data. The solid lines in Fig. 2 show fits to the high
temperature magnetisation data for $T>$ 50 K with the function, $\chi
_{mol}=\frac{C}{T-\Theta}+\chi^{0}.$ The Curie-Weiss term with $C=(1-x)N_{A}%
\mu_{eff}^{2}/3k_{B},$ Avogadro number, N$_{A}$ , effective moment, $\mu
_{eff}$ , and Boltzman constant, k$_{B}$, accounts for the paramagnetic
component. The \textit{T} independent term, $\chi^{0},$ accounts, besides
Van-Vleck paramagnetism and core diamagnetism, for the strong correlation
effects within the clusters as detailed in \cite{Daghofer06}. The parameters
obtained at $x$=0.97 for the field directions parallel ($\Vert$) and
perpendicular ($\bot$) to the c-axis are $\Theta_{\Vert,\bot}$=-61 and -45 K,
$\chi_{\Vert,\bot}^{0}$ = 1$\cdot10^{-4}$ and 0.78$\cdot10^{-4}$ cm$^{3}$/mol
and C$_{\Vert,\bot}$=8.4$\cdot10^{-3}$ and 5.5$\cdot10^{-3}$ cm$^{3}$/mol,
respectively. For 3\% doping this translates into a moment per doped hole of
$\mu_{eff}=1.5$ and $1.2$ in units of Bohr magneton, $\mu_{B},$ respectively.
With a spin only Land\'{e} factor of $g=2$ we thus derive a total spin of
S$_{\Vert,\bot}=0.4$ and $0.28$ per cluster. Such a small magnetic moment per
cluster needs to be reconciled with the sizeable moment per magnetic Co ion of
about 0.1-0.3 $\mu_{B}$ that has been inferred from $\mu$SR\ and neutron
experiments \cite{Bayrakci03,Mendels05,Bayrakci05,Helme05}. While this value
was obtained for AF\ samples with $x\approx0.82$, our $\mu$SR data indicate
that the Co moments at $x$=0.97 are of similar magnitude. The local field at
the muon site, B$_{\mu},$ as deduced from the TF-$\mu$SR depolarization rate
of $\ \lambda_{o}^{TF}\sim23$ $\mu s^{-1}$ at $x$=0.97 (assuming randomly
oriented moments) of
$<$%
B$_{\mu}$%
$>$%
=$\frac{\lambda_{o}^{TF}}{\gamma_{\mu}}\approx270\ G$ ($\gamma_{\mu
}=851.4\ MHz/T$ is the muon gyromagnetic ratio) compares indeed well to the
one obtained from the highest precession frequency of the antiferromagnetic
samples of $\nu_{\mu}=$3.3 MHz (see Fig. 3) with B$_{\mu}$=$\frac{2\pi\nu
_{\mu}}{\gamma_{\mu}}\approx240\ G$. It is now interesting to note that the
non magnetic regions experience noticeable magnetic stray fields which
apparently originate from the magnetic clusters since they also undergo a
freezing transition around 20 K. This is evident from the ZF- and LF-$\mu
$SR\ data in Figs 1b and 1c which show that the slowly relaxing component
exhibits a decoupling effect for LF=50 Oe at 15 K that is absent at 30 K. The
large magnitude of the decoupling effect at 15 K suggests that stray fields
essentially persist throughout the entire non-magnetic volume of the sample.
This result needs to be compared with the estimates based on dipolar field
calculations which indicate that the stray fields around antiferromagnetically
ordered clusters are extremely short ranged, i.e. they are noticeable only
over a few nanometers. Accordingly, our combined $\mu$SR and dc magnetisation
data are indeed suggestive of a nanoscopic coexistence of the magnetic and the
non magnetic regions.

This conclusion is supported by our finding that with increasing hole doping,
1-$x$, the magnetic clusters percolate and form extended magnetic patches
until a homogeneous magnetic state is achieved near $x$=0.78. This is shown in
Figure 3 which displays the ZF-$\mu$SR time spectra for $x$=0.92, 0.87 and
0.78. The insets show the distribution in frequency space as obtained with a
maximum entropy analysis. It is evident that all the 5 K spectra contain an
oscillatory component due to magnetic order at least in parts of the volume.
We used the function $P(t)=P(0)\,\underset{i}{\sum}\;A_{i}\;\cos(2\pi\nu_{\mu
}^{i}t)\;\exp(-\lambda^{i}t)$ to obtain the parameters for the precession
frequency, $\nu_{\mu}$, the relaxation rate, $\lambda$, and the relative
amplitudes, $A$, as shown in\ Fig. 4 together with previous data for
$x$=0.82(1) \cite{Bayrakci03}. Starting from a threshold of $x^{tr}\approx
$0.04, the amplitude of the oscillatory signal increases rapidly with hole
doping, and approaches 100 \% around $x$=0.78. The full oscillatory amplitude
of the 5K spectrum at $x$=0.78 also confirms that the magnetic field at the
muon site, $\overrightarrow{B}_{\mu},$\ is almost perpendicular to the muon
spin direction (which is tilted by 60 degree against the c-axis). Accordingly,
the $\mu$SR amplitudes should indeed be proportional to the corresponding
magnetic volume fractions \cite{Bayrakci03}. The c-axis orientation of
$\overrightarrow{B}_{\mu}$ furthermore suggests that the muons reside at
interstitial sites near the oxygen ions rather than on vacant Na sites where
$\overrightarrow{B}_{\mu}$ would have a larger component parallel to the
CoO$_{2}$ layers and be much smaller in magnitude as was previously noted
\cite{Bayrakci03,Mendels05}. While we do not attempt\ here a detailed
assignment of the muon sites, we notice that the signal at $\nu_{\mu}$=1.2 MHz
likely arises from the muons that are stopped near the boundaries of the
magnetic patches, since it has a large relaxation rate of about 3 $\mu$%
s$^{-1}$, it dominates at $x$=0.92 and 0.87 and is absent at $x$=0.78.
Correspondingly, we suggest that the signals near 2.5 and 3.0 MHz, which
become very narrow and further split at $x$=0.78, arise from muons that reside
within the order patches. Whether the distinction between these muon sites
arises from different structural (for example due to the Na neighbors) or
magnetic environments requires further investigation.

Next, we discuss the remarkable trend that the $\mu$SR\ spin fluctuation rate,
$\tau_{\mu},$ at T%
$>$%
T$_{N}$ is strongly enhanced for $x$=0.78 as compared to $x$=0.97. At $x$=0.97
slow spin fluctuations with 10$^{-6}$ s $>\tau_{\mu}>$10$^{-10}$ s persist to
at least 200 K while at $x$=0.78 the spin fluctuations become too fast to be
detected within the $\mu$SR time window of $\tau_{\mu}$%
$<$%
10$^{-10}$ s immediately above $T_{N}$=22 K. This is evident from Fig. 3c
where already the 25 K spectrum exhibits a Gaussian shape and a small
relaxation rate that is entirely due to the nuclear moments. This trend is
supported by the magnetization data for $x$=0.78 (Fig. 2) which exhibit a
broad maximum around 60 K that is characteristic for low-dimensional and/or
geometrically frustrated systems with strong (quantum)fluctuations. We
emphasize that this behavior once more is not compatible with a purely
chemical phase separation where the spin fluctuations should be enhanced due
to finite size effects for the isolated nanoscopic magnetic clusters at
$x$=0.97. Our data therefore highlight that the ordered magnetic state at
$x$=0.78 is not a conventional kind of A-type antiferromagnet that should not
be subject to such strong frustration effect (not even for a triangular
lattice). 

In summary, by $\mu$SR and dc magnetisation measurements we have explored the
magnetic properties of a series of highly Na doped Na$_{x}$CoO$_{2}$ single
crystals with  0.78(1)$\leq x\leq$0.97(1). Our data provide evidence for an
intrinsically inhomogeneous magnetic state with hole doping (Na vacancy)
induced magnetic clusters that percolate at 1-$x\gtrsim0.04$ until they yield
a bulk magnetic state near $x=$0.78. We observed a distince magnetic provide
evidence that the 

The SST may be induced by the doped holes which lower the crystal field
symmetry of the neighboring Co$^{3+}$ ions and thus give rise to a splitting
of the e$_{g}$ levels that reduces the energy difference to the highest
occupied t$_{2g}$ level. Accordingly, at low doping each hole induces a
magnetic cluster with a central low spin (LS, S=1/2) Co$^{4+}$ ion that is
surrounded by six intermediate spin (IS, S=1) Co$^{3+}$ ions
\cite{Bernhard04,Khaliullin05}. The SST\ model predicts indeed that a
homogeneous long-range ordered magnetic state occurs at 1-$x$=0.25
\cite{Bernhard04,Khaliullin05}. Explicit calculations\ have shown that while
the dominant interactions within the clusters are AF, a weak residual FM
in-plane interaction between the clusters gives rise to an A-type AF\ state
\cite{Daghofer06,Khaliullin05}. Notably, the SST model predicts that the
induced IS Co$^{3+}$ ions reside on a 2-d Kagom\'{e} lattice
\cite{Bernhard04,Khaliullin05} which is geometrically frustrated and thus
subject to strong quantum fluctuations. Furthermore, the SST model accounts
for the rapid disappearance of magnetism at 1-$x$%
$>$%
0.25 where the symmetry reduction associated with the doped holes becomes
insufficient to induce a SST. It also accounts for the evolution of the
magnetic volume fraction at low doping, although we note that this can be
equally well explained by chemical phase separation into non magnetic Na$_{1}%
$CoO$_{2}$ and magnetic Na$_{0.85}$CoO$_{2}$ \cite{Julien05}. At 1-$x$=0.03
the SST model predicts that each hole gives rise to a magnetic cluster which
contains 24 oxygen ions and subsequent muon sites \cite{Bayrakci03,Mendels05}
that are adjacent to at least one magnetic Co ion. Accordingly, due to the
O:Co ratio of 2:1, it yields a fraction of 36\% magnetic muon sites which
agrees well with our experimental value of about 40\%. The SST model also
allows one to understand the glassy behavior and the strongly reduced spin
fluctuation rate of the isolated clusters in terms of a delicate energy
balance between single-, double- or even triple hole clusters that likely
depends on the spatial arrangement of the Na vacancies. The related disorder
and the distribution in the shape anisotropy of the clusters thus may well
account for the glassy magnetic properties at 1-$x$=0.03.

We conclude by pointing out that a corresponding doping induced SST\ is indeed
well established for the simple perovskite La$_{1-x}$Sr$_{x}$CoO$_{3}$ whose
Co ions are in a corresponding valence state and crystallographic coordination
\cite{Potze95}. Nevertheless, the possible SST in Na$_{x}$Co$_{2}$ has some
unique aspects like the small total moment of the magnetic clusters of only
$\mu_{eff}\sim1.5\mu_{B}$ and the circumstance that the IS\ Co moments reside
on a Kagom\'{e} lattice and thus are geometrically frustrated. Both of them
favor strong quantum fluctuations which may well be relevant for the
unconventional magnetic, electronic and thermoelectric properties of these
fascinating materials.

We gratefully acknowledge the financial support of the Swiss Science
Foundation (SNF) through grant 200021-111690/1. We thank A. Amato for the
technical support at PSI and D. Argyriou for the neutron activation analysis
at Hahn Meitner Institut.

%

\begin{figure}
[ptb]
\begin{center}
\fbox{\includegraphics[
natheight=3.480000in,
natwidth=3.612300in,
height=3.5077in,
width=3.64in
]%
{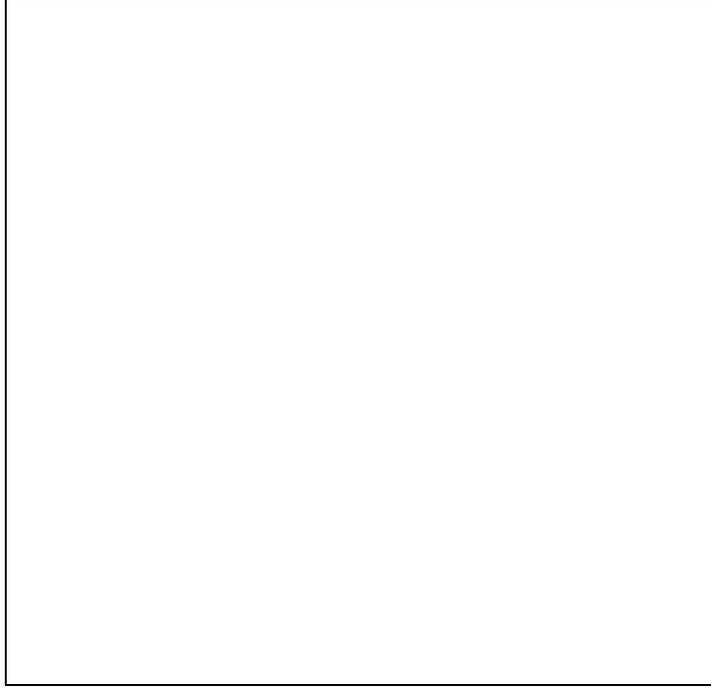}%
}\caption{The $\mu$SR data for Na$_{0.97}$CoO$_{2}$. (a) Displays the
zero-field spectra at representative temperatures. The solid lines show fits
as obtained with the function $P(t)/P(0)=0.4\cdot\exp\left(  -\lambda
_{1}t\right)  ^{\beta}+0.6\cdot KG\cdot\exp\left(  -\lambda_{2}t\right)  $
where KG represents the so-called Kubo-Toyabe function which accounts for the
nuclear magnetic moments. (b) and (c) show longitudinal-field spectra at 30 K
and 15 K, and (d) the transverse field (TF) relaxation rate of the fast
relaxing component. Inset: X-ray diffraction data (Cu$_{K\alpha}$ with
wavelength, $\lambda=$1.54 \AA ) for crystals from two different growth
batches with $x$=0.97(1) indicating the presence of a pure $\alpha$-phase or
mixed $\alpha$- and $\alpha$' phases, respectively. Arrows mark the position
of the $\alpha$'-phase (002) peak and the $\alpha$-phase (003) peaks.}%
\label{Figure 1}%
\end{center}
\end{figure}
\bigskip%

\begin{figure}
[ptb]
\begin{center}
\includegraphics[
trim=0.000000in 0.000000in -0.000491in 0.000000in,
natheight=2.348800in,
natwidth=2.453500in,
height=2.3756in,
width=2.4811in
]%
{Bernhard-Fig2.wmf}%
\caption{\textit{T}-dependent magnetic susceptibility of Na$_{x}$CoO$_{2}$
with $x$=0.97 and 0.78. Solid lines show Curie-Weiss fits for $x$=0.97 as
specified in the text. The arrow marks the freezing transition at $T_{f}$=20.7
K.}%
\label{Figure 2}%
\end{center}
\end{figure}
%

\begin{figure}
[ptb]
\begin{center}
\includegraphics[
natheight=2.743200in,
natwidth=2.932600in,
height=2.7709in,
width=2.9603in
]%
{Bernhard-Fig3.wmf}%
\caption{Zero-field $\mu$SR spectra for Na$_{x}$CoO$_{2}$ crystals with
$x$=0.92, 0.87 and 0.78 which exhibit a rapidly increasing fraction of the
oscillatory signal below $T_{N}$.}%
\label{Figure 3}%
\end{center}
\end{figure}
\bigskip%

\begin{figure}
[ptb]
\begin{center}
\includegraphics[
natheight=3.468800in,
natwidth=2.353100in,
height=3.4964in,
width=2.3808in
]%
{Bernhard-Fig4.wmf}%
\caption{Doping dependence of the fit parameters of the magnetic component in
the ZF-$\mu$SR spectra at 5\ K. (a) Amplitude of the precession frequencies.
The half down (up) filled symbols show the amplitudes of the splitted
frequencies at $x$=0.78 with 2.2 (2.6) MHz and 3.1 (3.3) MHz, respectively.
Solid (open) circles show the sum of the oscillatory (total magnetic)
amplitudes. The dotted line is a guide to the eye (b) Doping dependence of the
corresponding relaxation rates as shown by the same symbols.}%
\label{Figure 4}%
\end{center}
\end{figure}

\end{document}